\documentclass[11pt,twoside]{article}
\usepackage{CAGN2016}
\usepackage{graphicx}

\usepackage[T1]{fontenc} 

\usepackage{latexsym}
\usepackage{verbatim}

\begin{document}

\vskip 1.0cm
\markboth{J.~Garc\'{\i}a--Rojas}{Abundances in PNe from faint lines}
\pagestyle{myheadings}
%
%
\vspace*{0.5cm}
\parindent 0pt{Invited Review}


\vspace*{0.5cm}
\title{Chemical abundances in Galactic planetary nebulae from faint emission lines}

\author{J.~Garc\'{\i}a--Rojas$^{1,2}$}
\affil{$^1$Inst. de Astrof\'{\i}sica de Canarias, E-38200, La Laguna, Tenerife, Spain\\
$^2$Dept. de Astrof\'{\i}sica, Univ. de La Laguna, E-38206, La Laguna, Tenerife, Spain}

\begin{abstract}
Deep spectrophotometry has proved to be a fundamental tool to improve our knowledge on the chemical content of planetary nebulae. With the arrival of very efficient spectrographs installed in the largest ground-based telescopes, outstanding spectra have been obtained. These data are essential to constrain state-of-the-art nucleosynthesis models in asymptotic giant branch stars and, in general, to understand the  chemical evolution of our Galaxy. In this paper we review the last advances on the chemical composition of the ionized gas in planetary nebulae based on faint emission lines observed through very deep spectrophotometric data.
\end{abstract}

\section{Introduction}

Huggins \& Miller (1864) obtained the first spectrum of a planetary nebula (The Cat's Eye Nebula), where they detected a bright emission line coming from a misterous element that Huggins (1898) called ``nebulium''. Several decades after, Bowen (1927) showed that this emission was produced by doubly ionized oxygen (O$^{2+}$). The first ``deep'' spectra of planetary nebulae (hereinafter, PNe) were obtained by Wyse (1942), who detected for the first time very faint metal recombination lines of C~{\sc ii} and O~{\sc ii} in the spectrum of several PNe and in the Orion nebula. 

\begin{figure}  
\begin{center}
\hspace{0.25cm}
\includegraphics[width=\textwidth]{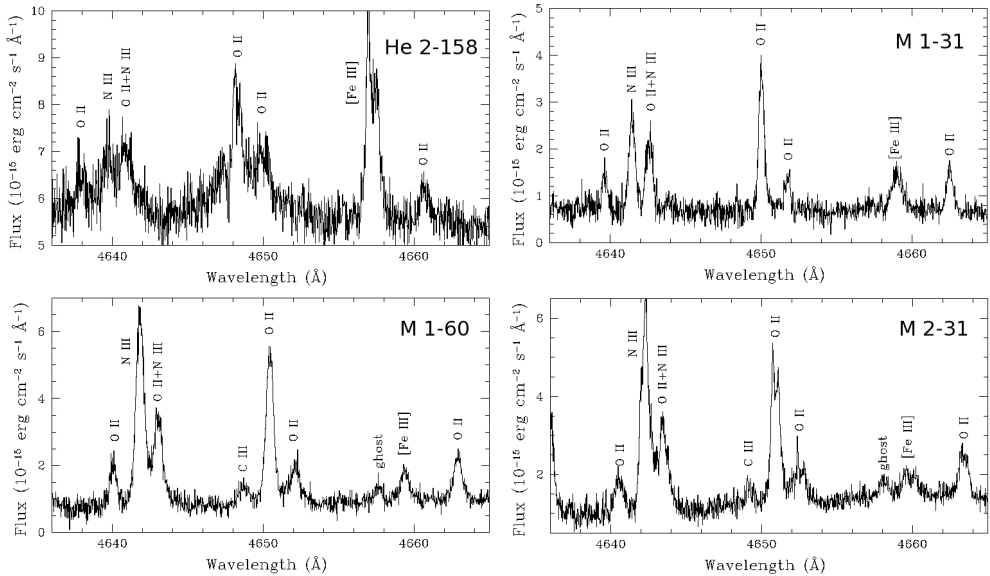}
\caption{Portion of the high-resolution (R$\sim$15000) spectra showing the zone where the multiplet 1 O~{\sc ii} lines lie. As it can be shown the quality of the spectra allows one to detect several other permitted lines of C, O, and N (Garc\'{\i}a--Rojas et al. in prep.),  to deblend several very close emission lines. }
\label{oii_lines}
\end{center}
\end{figure}

Since these early achievements in spectrophotometry of PNe, the amount of deep optical and near-infrared spectrophotometric data of PNe has increased significantly. This has been possible thanks to both the developement of more efficient instruments and of ground-based telescopes with large collecting areas. In Table~\ref{table1} we show some of the deepest optical spectra ever taken for Galactic PNe in the last 25 years with the purpose of obtaining chemical abundances. Although this list does not intend to be complete, it is quite representative of the last advances in the field. To progress in the field of chemical abundances in ionized nebulae we need to obtain very deep spectra to detect the weakest lines and, additionally, we need the spectra to be of high resolution to properly measure important faint emission lines which are blended with other features. As it can be seen in Table~\ref{table1}, the deepest spectra of PNe (those of NGC\,7027 and NGC\,7009) have been taken at relatively low-resolution. However, the authors of these works did a very careful work of deblending using ``ad-hoc'' atomic physics. On the other hand, high-resolution spectra can reveal the kinematical structure of the nebula, resulting in different profiles depending on the excitation of the emitting ion (see e.~g. Sharpee et al. 2003, Garc\'{\i}a--Rojas et al. 2015). This makes line identification easier and allows one to deblend important emission lines. Additionally, the distinct line profiles provide a valuable tool to identify the physical region where individual lines are formed (e.~g. Richer et al. 2013, Pe\~na et al. 2017). In Fig.~\ref{oii_lines} we show a portion of the deep, high-resolution (R$\sim$15000) spectra of several PNe, where it is clear that only at such a high-resolution the O~{\sc ii} lines can be properly deblended from other N and C emission features (see Sect.~\ref{adfs}). 

In the following sections, we discuss recent advances in chemical abundances in PNe from the analysis of very deep optical and near-infrared spectra. 

\setcounter{table}{0}
\begin{table}
\centering
\caption{Deep spectra of PNe.}
\label{table1}
\begin{tabular}{cccccc}
\noalign{\hrule} \noalign{\vskip3pt}
&  &  & Wavelength & Number &  \\
Reference$^{\rm a}$& Telescope & Objects & range (nm) & of lines &  R$\sim$$\lambda$/$\Delta\lambda$ \\
\noalign{\vskip3pt} \noalign{\hrule} \noalign{\vskip3pt}
1 & OHP 1.93m 	& NGC\,7027 	& 399--1050 	& $\sim$680 	& $<$4000  	 	\\
2 & Lick 3m 	& IC\,4997 	& 360--1005 	& $\sim$470 	& $\sim$30000   	\\
3 & ESO 1.52m 	& NGC\,6153 	& 304--743 	& $\sim$380 	& $<$4000  	 	\\
4 & CTO 4m 	& IC\,418 		& 350--987 	& $\sim$800	& $\sim$20000   	\\
5 & VLT 8.2m 	& NGC\,5315 	& 300--1040 	& $\sim$550 	& $\sim$9000  	 	\\
6 & WHT 4.2m  	& NGC\,7027  	& 331--916 	& $\sim$1170 	& $<$5000  	 	\\
7 & KPNO 4m 	& NGC\,7027 	& 460--920 	& $\sim$750 	& $\sim$15000   	\\
   & LCO 6.5m 	& 3 PNe 		& 328--758 	& $\sim$2330 	& 22000--28000  	\\
8 & WHT 4.2m 	& NGC\,7009  	& 304--1100 	& $\sim$1200 	&  $<$5000   	 	\\
   & ESO 1.52m 	&   		 	& 		      	&			&  		 	 	\\
9 & LCO 6.5m 	& 12 PNe 		& 335--940 	& $>$3000 	& 22000--28000  	\\
10 & VLT 8.2m  	& NGC\,3918 	& 300--1040  	& $\sim$750 	&  $\sim$40000  	\\
\noalign{\vskip3pt} \noalign{\hrule} \noalign{\vskip3pt}
\end{tabular}
\begin{description}
\item[$^{\rm a}$] (1) P\'equignot \& Baluteau (1994); (2) Hyung et al. (1994); (3) Liu et al. (2000); (4) Sharpee et al. (2004); (5) Peimbert et al. (2005); (6) Zhang et al. (2005); (7) Sharpee et al. (2007); (8) Fang \& Liu (2011); (9) Garc\'{\i}a--Rojas et al. (2012); (10) Garc\'{\i}a--Rojas et al. (2015).
\end{description}
\end{table}

\section{The abundance discrepancy problem}
\label{adfs}

The \textit{abundance discrepancy problem} is one of the major unresolved problems in nebular astrophysics and it has been around for more than seventy years (since Wyse 1942). It consists in the fact that in photoionized nebulae -- both H~{\sc ii} regions and planetary nebulae (PNe)-- optical recombination lines (ORLs) provide abundance values that are systematically larger than those obtained using collisionally excited lines (CELs).
Solving this problem has obvious implications for the measurement of the chemical content of nearby and distant galaxies, because this task is most often done using CELs from their ionized interstellar medium.

For a given ion, the abundance discrepancy factor (ADF) is defined as the ratio between the abundances obtained from ORLs and CELs, i.~e.,

\begin{equation}
{\rm ADF}({\rm X}^{i+}) = ({\rm X}^{i+}/{\rm H}^+)_{\rm ORLs}/({\rm X}^{i+}/{\rm H}^+)_{\rm CELs},
\label{adf}
\end{equation}

\noindent
and is usually between 1.5 and 3 in H~{\sc ii} regions and the bulk of PNe (see e.~g. Garc\'\i a--Rojas \& Esteban 2007, McNabb et al. 2013), but in PNe it has a significant tail extending to much larger values. 

The reason for this discrepancy has been discussed for many years and three main scenarios have been proposed: i) the existence of temperature fluctuations over the observed volume of the nebula (Peimbert 1967, Torres--Peimbert et al. 1980), b) the presence of cold and dense H-poor droplets as the origin of the bulk of the ORL emission (e.~g. Liu et al. 2000, Stasi\'nska et al. 2007, Henney \& Stasi\'nska 2010), c) the departure of the free electron energy distribution from the Maxwellian distribution ($\kappa$-distribution, see Nicholls et al. 2012). However, in the last years, several works have argued against the $\kappa$-distribution as being esponsible for the abundance discrepancy in PNe (see Mendoza \& Bautista 2014, Storey \& Sochi 2014, Ferland et al. 2016). Unfortunately, there are so far not direct observational evidences that can favour any of these scenarios.

In the last 20 years the group led by Prof. X. --W. Liu (U. Beijing) have developed deep medium-resolution spectrophotometry of dozens of PNe (and high-resolution of a few PNe, see McNabb et al. 2016) to compute the physical and chemical properties of these objects from ORLs. 
In one of the most detailed and comprehensive studies of this group, Wang \& Liu (2007) showed that the values of the ADF deduced for the four most abundant second-row heavy elements (C, N, O and Ne) are comparable (see their Fig.~18). However, they also computed abundances from ORLs from a third-row element (Mg) and they found that the ORL abundance of magnesium is compatible with the solar photospheric value (even taking into account the small depletion expected for this element onto dust grains - less than 30\%).
Finally, these authors also showed that, regardless of the value of the ADF, both CEL and ORL abundances yield similar relative abundance ratios of heavy elements such as C/O, N/O and Ne/O . This has important implications, especially in the case of the C/O ratio, given the difficulties of obtaining this ratio from CELs (see Sect.~\ref{co_ratio}).

The Liu's group have strongly argued in favour of the inhomogeneus composition of  PNe and against pure temperature fluctuations. Some of their arguments are the following: i) far-IR [O~{\sc iii}] CELs, which in principle, have a much lower dependence on electron temperature than optical CELs, provide abundances that are consistent with those derived from optical CELs (see e.~g. Liu et al. 2000); ii) the analysis of the physical conditions using H, He, O and N ORLs yields electron temperatures that are much lower than those computed from classical CEL diagnostic ratios (see Zhang et al. 2004, 2009, Tsamis et al. 2004, Fang et al. 2011); additionally, ORL density diagnostics provide densities that are higher than those derived from CEL diagnostics; iii) chemically homogeneous photoionization models do not reproduce the required temperature fluctuations to match CEL and ORL abundances, while bi-abundance photoionization models including an H-poor (i.~e. metal-rich) component of the gas successfully reproduce the observed intensities of both CELs and ORLs (e.~g. Yuan et al. 2011). All these arguments strongly favour the presence of a low-mass component of the gas that is much colder and denser than the ``normal'' gas, and that is responsible for the bulk of the ORL emission. 
However, we cannot rule out the possibility that different physical phenomena can contribute simultaneously to the abundance discrepancy in PNe.

Some physical phenomena have been proposed to explain the abundance discrepancy in the framework of temperature fluctuations or chemical inhomogeneities scenarios (Peimbert \& Peimbert 2006, Liu 2006). 
Some recent works on the Orion nebula have observationally linked the abundance discrepancy to the presence of high velocity flows (Mesa-Delgado et al. 2009) or to the presence of high density clumps, such as protoplanetary disks (Mesa--Delgado et al 2012, Tsamis et al. 2011). On the other hand, Liu et al. (2006) found a very extreme value of the ADF for the PN Hf\,2--2 (ADF$\sim$70) and, for the first time, speculated with the possibility that this large ADF could be related to the fact that the central star of the PN, which is a close-binary star, has gone through a common-envelope phase. 

By means of spectroscopic observations at the 4.2m William Herschel Telescope (WHT) in La Palma, Spain, Corradi et al. 2015 recently confirmed the hipothesis proposed by Liu et al. (2006) that the largest abundance discrepancies are reached in PNe with close-binary central stars, the most extreme object being the PN Abell~46, where we have found an ADF(O$^{2+}$)$\sim$120, and as high as 300 in its inner regions. Their spectroscopic analysis supports the previous interpretation that, in addition to ``standard'' hot (T$_e$$\sim$10$^4$~K) gas, a colder (T$_e$$\sim$10$^3$~K), metal-rich, ionized component also exists in these nebulae.  Both the origin of the metal-rich component and how the two gas phases are mixed in the nebulae are basically unknown. Moreover, this dual nature is not predicted by mass-loss theories. However, it seems clear that the large-ADF phenomena in PNe is linked to the presence of a close-binary central star. In fact, Wesson et al. (2017) recently completed a survey of the ADFs in seven PNe with known close-binary central stars and they found ADFs larger than 10 for all of them, confirming the strong link between large ADFs and close-binary central stars.
On the other hand, several spectroscopic studies have shown that the ORL emitting plasma is generally concentrated in the central parts of the PNe. This occurs in PNe with known close-binary central stars and large ADFs (e.~g. Corradi et al. 2015, Jones et al. 2016), in PNe with low-to-moderate ADFs and no indication of binarity (e.~g. Liu et al. 2001, Garnett \& Dinerstein 2001) and in PNe with relatively large ADFs but no known close-binary central star (e.~g. M\,1--42, see Climent 2016, Garc\'{\i}a--Rojas et al. 2017).

\begin{figure}  
\begin{center}
\hspace{0.25cm}
\includegraphics[width=\textwidth]{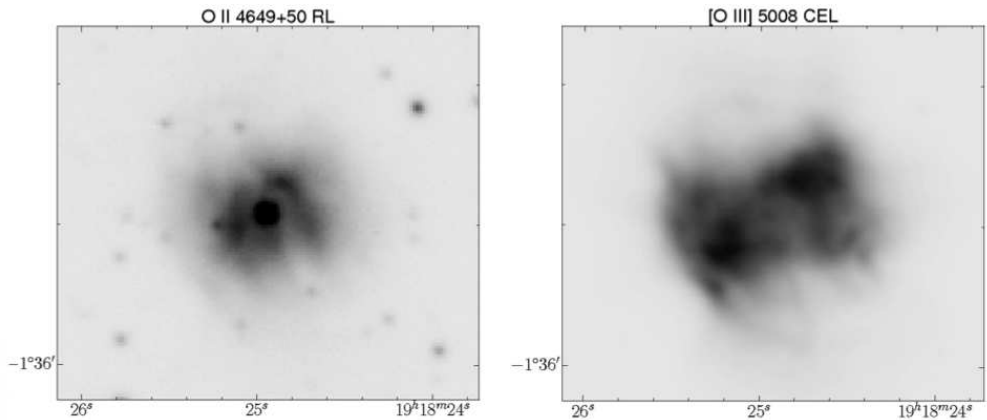}
\caption{OSIRIS-GTC tunable filter image of NGC~6778 in the O~{\sc ii} $\lambda\lambda$4659+51 ORLs (left) and  ALFOSC-NOT image of Guerrero \& Miranda (2012) in the [O~{\sc iii}] $\lambda$5007 CEL (right). Figure adapted from figure 2 of Garc\'{\i}a-Rojas et al. (2016).}
\label{tf_orls}
\end{center}
\end{figure}

Garc\'{\i}a--Rojas et al. (2016) recently obtained the first direct image of the PN NGC\,6778 (a PN with ADF$\sim$20) in O~{\sc ii} recombination lines, taking advantage of the tunable filters available at the OSIRIS instrument in the 10.4m Gran Telescopio Canarias (GTC). They found that in NGC~6778, the spatial distribution of the O~{\sc ii} $\lambda$$\lambda$4649+50 ORL emission does not match that of the [O~{\sc iii}] $\lambda$5007 CEL (see Fig.~\ref{tf_orls}). Garc\'{\i}a--Rojas et al. (2017) found the same behaviour in Abell~46 using direct tunable filter images centred at $\lambda\lambda$4649+51 \AA.

Moreover, Garc\'{\i}a--Rojas et al. (2017) obtained deep 2D spectroscopic observations with MUSE at the 8.2m Very Large Telescope (VLT) of five southern large-ADF PNe, and they confirmed this behaviour in at least the PNe Hf~2-2 (ADF$\sim$84), M~1-42 and NGC\,6778 (both with ADF$\sim$20).

These results clearly support the hypothesis of the existence of two separate plasmas, at least in these large-ADF PNe, with the additional indication that they are not well mixed, perhaps because they were produced in distinct ejection events related to the binary nature of the PN central star. Wesson et al. (2017)  propose that a nova-like outburst from the close-binary central star could be responsible for ejecting H-deficient material into the nebulae soon after the formation of the main nebula.

\section{The C/O ratio from recombination lines}
\label{co_ratio}

The determination of accurate C/H and C/O ratios in H~{\sc ii} regions is of paramount importance to constrain chemical evolution models of galaxies. Moreover, C/O determinations in PNe can also constrain nucleosynthesis processes in low-to-intermediate mass stars. However, the determination of C abundances in photoionized nebulae is difficult because, traditionally, C abundances have been derived from the semi-forbidden C~{\sc iii}] $\lambda$1909 and C~{\sc ii}] $\lambda$2326 CELs in the UV, which can only be observed from space and whose intensity is strongly affected by interstellar reddening and extremely dependent on the electron temperature.

Alternatively, there is a method to derive C abundances in ionized gas based on the faint C~{\sc ii} $\lambda$4267 RL. Thanks to the new CCDs with improved efficiency in the blue and the use of large telescopes, several high-quality observations of the C~{\sc ii} $\lambda$4267 RL in PNe have been achieved in the last years (e.~g.  Peimbert et al. 2004, Liu et al. 2004b, Wesson et al. 2005, Sharpee et al. 2007, Wang \& Liu 2007, Garc\'{\i}a--Rojas et al. 2009, 2013, Fang \& Liu 2013). 

C/O ratios derived from RLs combined with other abundance ratios as N/O or He/H, can set strong constraints to the initial mass of PNe progenitors. This is owing to different processes occuring at the interior of AGB stars (third dredge-up episodes, hot bottom burning process) activates at different masses and can strongly modify C/O and N/O ratios (see e.~g. Karakas \& Lugaro 2016 and references therein). Additionally, C/O ratios can be used to obtain information about the efficiency of dust formation in C-rich or O-rich environments (see below) and to learn about different dust-formation mechanisms (see Garc\'{\i}a--Hern\'andez et al. 2016).


\section{Faint emission lines of refractory elements}
\label{fe_ni}

Iron emission lines are relatively faint in the spectra of Galactic PNe. Delgado--Inglada et al. (2009) computed detailed Fe abundances for a sample of 28 PNe and found that more than 90\% of Fe atoms are condensed on dust grains. These authors did not find differences between the iron abundances in C-rich and O-rich PNe, suggesting similar depletion efficiencies in both environments. 

Delgado--Inglada \& Rodr\'{\i}guez (2014) combined C/O ratios derived from both UV CELs and optical ORLs (as we comment in Sect.~\ref{adfs}, they seem to be equivalent) with information obtained from \textit{Spitzer} mid-infrared spectra. They also computed Fe depletions, and found that the highest depletion factors were found in C-rich objects with SiC or the 30 $\mu$m feature in their infrared spectra, while the lowest depletion factors were found for some of the O-rich objects showing silicates in their infrared spectra.

Delgado--Inglada et al. (2016) compiled detections of very faint [Ni~{\sc ii}] and [Ni~{\sc iii}] lines in deep spectra of Galactic PNe and H~{\sc ii} regions. They determined the nickel abundance from the [Ni~{\sc iii}] lines using an extensive grid of photoionization models to determine a reliable ionization correction factor (ICF). From the comparison of Fe/Ni ratios with the depletion factor obtained from both [Fe/H] and [Ni/H], they conclude that nickel atoms are more efficiently stuck to dust grains than iron atoms in environments where dust formation or growth is more important.

\section{Neutron-capture element abundances in PNe}
\label{s_elements}

Nebular spectroscopy of neutron(\textit{n})-capture elements (atomic number Z $>$ 30) is a recent field that has seen rapid development in the last 10 years, and holds promise to significantly advance our understanding of AGB \textit{n}-capture nucleosynthesis. Nebular spectroscopy 
can reveal unique and complementary information to stellar spectroscopy. 
Observations of PNe provide the first opportunity to study the production of the lightest \textit{n}-capture elements (Z $\le$ 36) and noble gases (Kr and Xe) in one of their sites of origin. Unlike the case of AGB stars, nucleosynthesis and convective dredge-up are complete in PNe, whose envelopes contain material from the last 2$-$3 thermal pulses. 
Accurate computations of \textit{n}-capture elements would shed light on the different scenarios proposed  for the production of these elements and would constrain the chemical yields of low- and intermediate-mass stars for these elements. 

\textit{n}-capture elements were not recognized in any astrophysical nebula until P{\'e}quignot \& Baluteau (1994) identified emission lines of Br, Kr, Rb, Xe, Ba, and possibly other heavy species in the bright PN NGC\,7027. Since then, a breathtaking number of \textit{n}-capture element emission lines have been identified for the first time in near-infrared (Dinerstein 2001, Sterling et al. 2016, Sterling et al. 2017), UV (Sterling et al. 2002) and optical (Sharpee et al. 2007, Garc\'{\i}a--Rojas et al. 2012, 2015) spectra of PNe. The new detections have led to a dedicated effort to produce atomic data needed for abundance determinations (e.~g., see Sterling \& Witthoeft 2011, Sterling et al. 2016, and references therein). 
The new photoinization cross-sections and recombination coefficients have been incorporated in photoionization calculations to compute reliable ICFs (e.~g., Sterling et al. 2015). The new collisional strenghts have been used for abundance determinations of newly detected ions (see Sterling et al. 2016, for [Rb~{\sc iv}], Sterling et al. 2017, for [Se~{\sc iii}] and [Kr~{\sc vi}]). 
Thanks to the fast advances in observations, atomic data determinations and numerical modeling, this field has grown up from just 3 PNe with \textit{n}-capture element abundances in 2001 to more than 100 Galactic PNe in 2016 (Sharpee et al. 2007, Sterling \& Dinerstein 2008, Sterling et al. 2016, Garc\'{i}a-Rojas et al. 2012, 2015).

The importance of deep, high-resolution optical spectrophotometry of PNe to detect faint \textit{n}-capture elements can be understood when comparing the works by Sharpee et al. (2007) and Garc\'{\i}a--Rojas et al. (2015). In the first case, several \textit{n}-capture emission lines were discovered in the spectra of 5 PNe, but even at a resolution of $\sim$22000, many features were not unambiguosly detected. Garc\'{\i}a--Rojas et al. (2015) took advantage of the very high-resolution (R$\sim$40000) spectrum of NGC\,3918 to clearly identify several ions of Kr, Xe, Rb and Se, testing for the first time the complete set of ICFs for Kr created by Sterling et al. (2015). Finally, Madonna et al. (2017) have combined a deep optical spectrum and a near-infrared spectrum of NGC\,5315, testing for the first time the complete set of ICFs for Se created by Sterling et al. (2015). 

\section{Other faint emission lines in optical and IR spectra of PNe}
\label{other}

Other faint emission lines have been detected in the optical and mid-IR spectra of PNe and are of great interest for different reasons. I am going to focus on F, Zn and $\alpha$--elements Ca, K and Na. 

Zinc is a useful surrogate element for measuring abundances of the iron group because, unlike iron, it is not depleted onto dust grains. Smith et al. (2014) developed an observational campaign with ISAAC at the VLT in the mid-IR range with the aim of detecting Zn emission lines. These authors successfully detected the faint [Zn~{\sc iv}]  at 3.625 $\mu$m in seven objects and added other two detections observed by Dinerstein \& Geballe (2001). They compute Zn elemental abundances using photoionization models and conclude that assuming Zn/Zn$^{3+}$$\sim$O/O$^{2+}$ is a good approximation. The main conclusion of their work was that the majority of the sample exhibit subsolar [Zn/H], and half of the sample show enhacement in [O/Zn].

Fluorine faint emission lines have been detected and measured in deep optical spectra of PNe. This element is very interesting because it only has one stable isotope, $^{19}$F, and several possible nucleosynthetic origins in the literature have been proposed: i) explosions of Type II supernovae, ii) stellar winds from massive Wolf-Rayet stars, and iii) the occurence of the third dredge-up in AGB stars (see Zhang \& Liu 2005). Accurate abundance determinations of fluorine can shed some light on its nucleosynthetic origin. 
Zhang \& Liu (2005) performed a survey in the literature looking for detections of the faint [F~{\sc ii}] $\lambda$4789 and [F~{\sc iv}] $\lambda$4060  lines. To compute elemental F abundances, these authors assumed that F/O=F$^+$/O$^+$ and F/O=(F$^{3+}$/Ne$^{3+}$)(Ne/O) for low-- and high--ionization PNe, respectively. They found that F is generally overabundanct in PNe and, that F/O is positively correlated with C/O, which favours the hypothesis of fluorine being produced in thermally pulsing AGB stars. On the other hand, Otsuka et al. (2008) detected the faint [F~{\sc iv}] $\lambda\lambda$3997, 4060 lines in the extremely metal-poor halo PN BoBn\,1. They found a strong enhancement of the F/H ratio and by comparing the abundance pattern of  BoBn\,1 with carbon-enhanced metal-poor stars abundances, they concluded that the observed behaviour could be explained if the nebula evolved from a binary system.

Finally, Garc\'{\i}a--Rojas et al. (2015) reported, for the first time, the detection of faint sodium lines ([Na~{\sc iv}] $\lambda$3242 \AA  and $\lambda$3362 \AA) in the deep spectrum of NGC\,3918. These detections along with the [Na~{\sc iii}] $\lambda$7.31 $\mu$m by Pottasch et al. (2009) can be used to constrain a grid of photoionization models computed by Delgado--Inglada et al. (2014) to construct, for the first time, an ICF to estimate the total Na abundance (Medina--Amayo et al. in prep.). A similar study can be developed taking advantage of the few detections of potasium and calcium lines in the spectra of PNe.

\acknowledgments I deeply thank Oli Dors Jr. for inviting me to give this review and for giving me financial support to attend to the workshop. I am indebted to C. Esteban, C. Morisset and G. Stasi\'nska for discussions and a critical reading of this manuscript. I also thank all my collaborators and students
for many discussions on the research topics presented in this paper.

\end{document}